# Strain Engineering of Selective Chemical Adsorption on Monolayer MoS$_2$

Liangzhi Kou[1*], Aijun Du[2], Changfeng Chen[3] and Thomas Frauenheim[1]

1. Bremen Center for Computational Materials Science, University of Bremen, Am Falturm 1, 28359 Bremen, Germany
2. School of Chemistry, Physics and Mechanical Engineering, Queensland University of Technology, Brisbane, QLD 4001, Australia
3. Department of Physics and Astronomy and High Pressure Science and Engineering Center, University of Nevada, Las Vegas, Nevada, 89154, United States

*Corresponding author: kouliangzhi@gmail.com

Nanomaterials are prone to influence by chemical adsorption because of their large surface to volume ratios. This enables sensitive detection of adsorbed chemical species which, in turn, can tune the property of the host material. Recent studies discovered that single and multi-layer molybdenum disulfide (MoS$_2$) films are ultra-sensitive to several important environmental molecules. Here we report new findings from ab inito calculations that reveal substantially enhanced adsorption of NO and NH$_3$ on strained monolayer MoS$_2$ with significant impact on the properties of the adsorbates and the MoS$_2$ layer. The magnetic moment of adsorbed NO can be tuned between 0 and 1 $\mu_B$; strain also induces an electronic phase transition between half-metal and metal. Adsorption of NH$_3$ weakens the MoS$_2$ layer considerably, which explains the large discrepancy between the experimentally measured strength and breaking strain of MoS$_2$ films and previous theoretical predictions. On the other hand, adsorption of NO$_2$, CO, and CO$_2$ is insensitive to the strain condition in the MoS$_2$ layer. This contrasting behavior allows sensitive strain engineering of selective chemical adsorption on MoS$_2$ with effective tuning of mechanical, electronic, and magnetic properties. These results suggest new design strategies for constructing MoS$_2$-based ultrahigh-sensitivity nanoscale sensors and electromechanical devices.

Molybdenum disulfide ($MoS_2$) nanosheets have recently attracted great interest for their intriguing properties that promise a viable alternative to graphene-based systems.[1-4] Recent research indicated that single and multi-layer $MoS_2$ films are ultra-sensitive to a number of molecules important in environmental studies.[5-8] In additional, the semiconducting nature of $MoS_2$ is of great interest for modulating the transport characteristics with exposure to light or gate bias to enhance sensing performance.[9] Initial results have revealed that $MoS_2$ films are extremely sensitive to adsorption of $NO_x$ and $NH_3$[9-11] as indicated by changes in resistivity induced by adsorbed molecules that act as charge donors or acceptors. The detection of $NH_3$ is sensitive to 1 ppb (parts per billion), and the ultimate sensitivity of single-molecule adsorption was suggested for $NO_2$[9]. These reports have raised promising prospects of developing new $MoS_2$-based nanodevices. An in-depth understanding of chemical adsorption on $MoS_2$ requires further exploration of the adsorption process and characteristics in response to external influence that may lead to improved performance or even new phenomena. It has been shown that charge control can effectively tune the $CO_2$ capture on boron nitride nanomaterials.[12] Recent experiments and theoretical calculations also demonstrated that the chemical reactivity and reaction rate of aryl diazonium molecules on graphene can be dramatically increased by mechanical strain.[13] These studies suggest that chemical adsorption on $MoS_2$ is likely to be susceptible to external influence, thus opening a new avenue to control adsorption process and tune material properties. In fact, strain effect is expected to have a particularly strong influence on $MoS_2$ since it undergoes a strain induced semiconductor-to-metal transition with considerable charge redistribution[14-17], a phenomenon absent in graphene.

Flexible electronics is another area of potential application for monolayer and multi-layer $MoS_2$ films that exhibit superior mechanical properties[18-20]. There is, however, conflicting reports on the strength and breaking strain of $MoS_2$. A recent experimental study[18] found that the effective Young's modules of monolayer $MoS_2$ is 22±4 Gpa with an effective breaking strain of 6~10%. In stark contrast, the critical strain is estimated to be about 20% from first-principles calculations based on density functional theory[14,21-22]. This large disparity between the experimental results and theoretical predictions has been attributed to the defect effect that is not considered in theoretical modeling and calculations. While defects usually tend to reduce breaking strain and material strength, it remains unclear to what extent they are responsible for the observed strength reduction in $MoS_2$. Meanwhile, chemical adsorption is known to play a prominent role in affecting mechanical properties of nanostructures. Unveiling the effect of gas molecule adsorption on mechanical properties of monolayer $MoS_2$ would provide insights for understanding this issue that is important to its application in flexible electronic nanodevices.

In this work, we show by ab initio calculations that a biaxial strain can effectively tune the adsorption strength of NO and $NH_3$ on monolayer $MoS_2$, producing significant changes in the mechanical, electronic and magnetic properties of the adsorbates and the $MoS_2$ layer. This strain-dependent adsorption behavior is driven by a marked increase in charge transfer between the gas molecules and the strained $MoS_2$ layer. As a result, NO-adsorbed monolayer $MoS_2$ enters a new half-metallic state prior to its strain-induced metallization, and the magnetic moment on NO is tunable by strain in the range of $0~1\mu_B$. With the adsorption of $NH_3$ molecules, the critical breaking strain of monolayer $MoS_2$ is reduced to 12.5%, which is much smaller than the previous theoretical prediction of 20% and more consistent with the experimental results. Our results also show that adsorption of $CO_2$, CO, or $NO_2$ on $MoS_2$ is insensitive to the strain condition in the $MoS_2$ layer, allowing selective discrimination among the adsorbed chemical species. Recent experimental manipulation of strained $MoS_2$ nanosheets[18-20] indicates it is feasible to apply strain engineering to control and distinguish chemical adsorption on monolayer $MoS_2$, which is also useful for constructing $MoS_2$-based nanodevices.

## Results and Discussion

**Selective strain-sensitive adsorption and structural characteristics** We have examined adsorption of NO, $NO_2$, CO, $CO_2$ and $NH_3$ on the surface of monolayer $MoS_2$ with a focus on the strain response of the adsorption, and have compared the results to experimental measurements and theoretical calculations. We first consider two strained states (0% and 10%) to compare the response of the adsorption process. We list in Table 1 key parameters, including the adsorption distance (see definition in Figure 1a), adsorption strength (defined as $E_{abs}=E_{tot}-E_{MoS2}-E_{gas}$, where $E_{tot}$, $E_{MoS2}$, $E_{gas}$ are total energy of gas adsorbed $MoS_2$, energies of gas molecule and $MoS_2$, respectively) and magnetic moment. While adsorption of NO and $NH_3$ are strain sensitive, other gas molecules show no appreciable response to applied strain. We therefore focus our discussions below on the cases of NO and $NH_3$ adsorption on $MoS_2$.

The fully relaxed structural configurations are presented in Fig. 1, where one can see that the gas molecules prefer to locate in the hexagonal center (Fig. 1a) and would move there even when initially placed on top of Mo or S atoms. For NO adsorption, the nitrogen (oxygen) atom is aligned toward (away from) the $MoS_2$ layer and forming an angle of 62 degree relative to the surface of the $MoS_2$ layer. The hydrogen atoms in ammonia point toward the three Mo atoms of one $MoS_2$ hexagonal ring while the nitrogen atom sits at the center, as shown in Fig. 1a and 1b. We notice that our calculated adsorption distance is smaller and adsorption strength generally larger than the recently reported values[23]. This is likely due to the less stringent structural determination procedure employed in the earlier work that did not perform an extensive search for the most stable adsorption configuration. We have cross checked our optimized structures obtained from SIESTA calculations using the VASP[25] code with the van der Waals interaction. The two sets of results are in excellent agreement, providing strong support for their validity.

Meanwhile, results in Table 1 show that adsorption of N-based gas molecules has lower adsorption energies (i.e., more negative, thus stronger binding) compared those for adsorption of CO and $CO_2$, which explains the higher sensitivity to $NO_x$ and $NH_3$ adsorption as seen in recent experiments[9-11]. The adsorption distance between NO and strain-free $MoS_2$ is 2.1 Å, but is reduced to 1.62 Å (a 23% reduction) at 10% strain. For ammonia, the distance reduction under the same strain condition is 14%, which is smaller but still significant. As a result of the distance reduction, the binding becomes stronger as evidenced by the larger values of adsorption strength (see Table 1). The strain response of other gas adsorptions, such as $NO_2$, CO and $CO_2$, is much smaller, indicating sensitive selectivity of strain enhanced gas adsorption on monolayer $MoS_2$.

**Table 1.** Calculated adsorption distance $d$ between gas molecules (including NO, $NO_2$, CO, $CO_2$ and $NH_3$) and the $MoS_2$ layer, adsorption energy ($E_{abs}$) and magnetic moment (M) when the $MoS_2$ layer is in equilibrium condition (0% strain) and under a 10% strain.

|  | NO | | $NO_2$ | | CO | | $CO_2$ | | $NH_3$ | |
| --- | --- | --- | --- | --- | --- | --- | --- | --- | --- | --- |
|  | 0% | 10% | 0% | 10% | 0% | 10% | 0% | 10% | 0% | 10% |
| $d$ (Å) | 2.1 | 1.62 | 2.15 | 2.15 | 2.45 | 2.39 | 2.74 | 2.56 | 2.79 | 2.39 |
| $E_{abs}$ (eV) | -0.54 | -0.72 | -0.94 | -0.91 | -0.42 | -0.43 | -0.45 | -0.47 | -0.49 | -0.55 |
| M ($\mu_B$) | 1.0 | 0.71 | 1.0 | 0.97 | ~ | ~ | ~ | ~ | ~ | ~ |

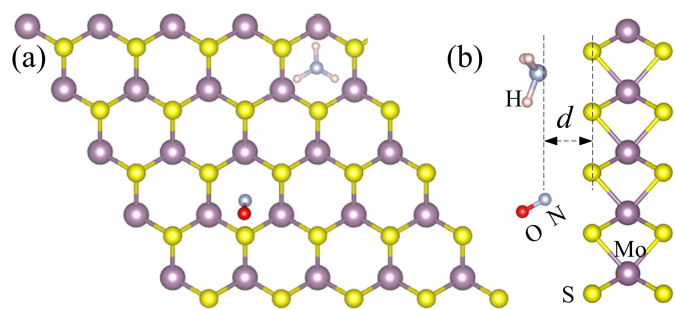

**Figure 1**. Top (a) and side (b) view of the relaxed structural (supercell) model of gas adsorption on monolayer $MoS_2$.

**Strain-dependent NO adsorption on monolayer MoS$_2$** Our calculations indicate that adsorption of NO on monolayer MoS$_2$ is highly sensitive to the strain condition in the layer. In Fig. 2(a), we present the adsorption distance and energy as a function of strain. Both quantities decrease with rising strain; the adsorption energy becomes more negative, indicating stronger binding at reduced adsorption distance. These quantities, however, do not change linearly with strain. There is a sharp drop around the strain of 10% for both adsorption distance and energy. This is correlated with the metallization of MoS$_2$ that occurs at a critical strain of about 10% as reported in previous studies[14-17]; the electron redistribution associated with the metallization has a significant effect on the adsorption of gas molecules. Recent experimental and theoretical investigations indicate that charge transfer is the reason for the decrease in the resistance of MoS$_2$,[9] and the enhancement of gas adsorption found here can also be attributed to the increased electron transfer after the metallization of the MoS$_2$ layer. This can be seen from our calculated results on the enhanced charge transfer ($\Delta\rho=\rho_{tot}-\rho_{MoS2}-\rho_{gas}$) in the 10% strained MoS$_2$ as shown in the insets of Fig. 2(a). After the metallization of the MoS$_2$ layer, some electrons are even transferred to Mo atoms and become slightly delocalized. In contrast, the charge transfer in the strain-free layer is limited between the gas molecule and the most adjacent S atom.

It has been found in graphene and MoS$_2$[23,25] that adsorption of paramagnetic molecules such as NO and NO$_2$ can lead to their spin polarization. Our calculated magnetic moment for NO on MoS$_2$ is 1 $\mu_B$, which is in agreement with previously reported results,[23,25] and the spatial spin distribution analysis shows that all the spin-unpaired electrons are localized on the NO molecule (see the inset of Fig. 2b). Interestingly, we find that the magnetic moment of the adsorbed NO molecule is tunable by strain in the MoS$_2$ layer, and its size decreases gradually with increasing strain up to 10% and then undergoes a sharp downward turn toward zero. A

Mulliken population analysis indicates that in the absence of applied strain there is 0.051e of up-spin electron transferred to MoS$_2$ while it is 0.034e for the down-spin electron. At 10% strain, the corresponding values are 0.151e and 0.102e. The strain dependence of the transfer of spin polarized electrons leads to the change of magnetic moment discovered in our study. It is noted that the spin polarized electrons are always located on the NO molecule regardless of the strain state (see the spin distribution shown in the inset of Fig. 2b).

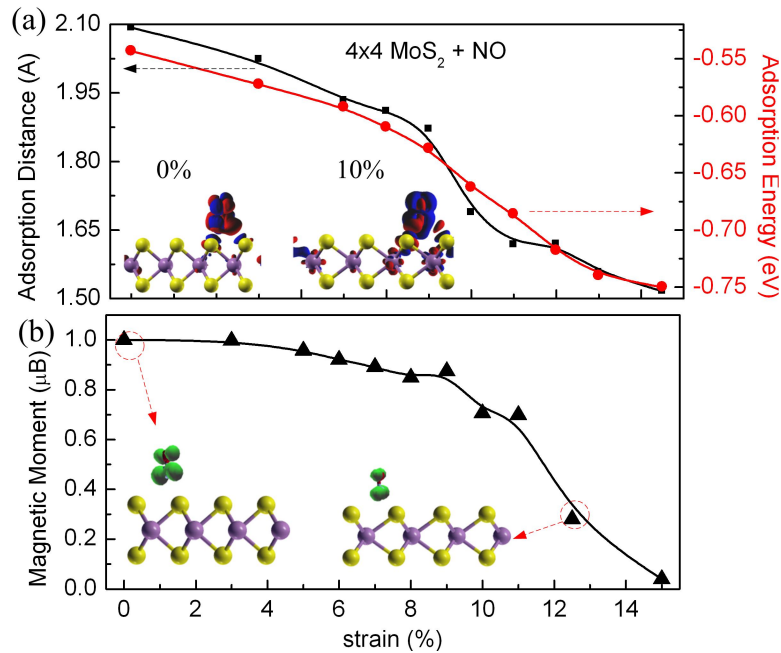

**Figure 2.** (a) Calculated adsorption distance and energy as a function of strain for NO adsorbed on monolayer MoS$_2$. The insets show electron transfer (i.e., the difference between electron distribution of NO+MoS$_2$ and that of separated components) at 0% and 10% strained conditions. The isosurface is $1\times10^{-3}$e/Å$^3$, and red (blue) indicates electron gain (loss). (b) Magnetic moment variation under strain, the spatial spin distributions at 0% and 12.5% strain are shown as insets.

The spin polarized density of states (DOS) of NO adsorbed monolayer MoS$_2$ is shown in Fig. 3, where results for three typical strained states (0%, 5% and 10%) are presented. The Fermi level after adsorption is shifted to the conduction band; it can be identified by the DOS plot (Fig.

3a) that indicates an n-type doping effect where NO acts as a charge donor. This effect is also manifested in the charge transfer from NO to MoS$_2$ as shown in the inset of Fig. 2a. The difference in the DOS near the Fermi level is mainly attributed to the N-p orbitals according to a projected DOS analysis, and the resulted DOS peak comes from only the spin-up electrons. This half-metallic state remains robust as strain increases (see the DOS of 5% strained NO+MoS$_2$ in Fig. 3b), and it lasts until the metallization of the MoS$_2$ layer at 10% strain. Figure 3c shows the DOS at 12.5% strain where the system is in a normal metallic state. The contribution of the DOS from the MoS$_2$ layer, which has a spin polarization opposite to that of the adsorbed NO, compensates the spin-polarized DOS from NO and eventually overwhelms it, leading to the normal metallic state. This explains the substantial decrease of the magnetic moment, especially after the metallization of the MoS$_2$ layer, as shown in Fig. 2b.

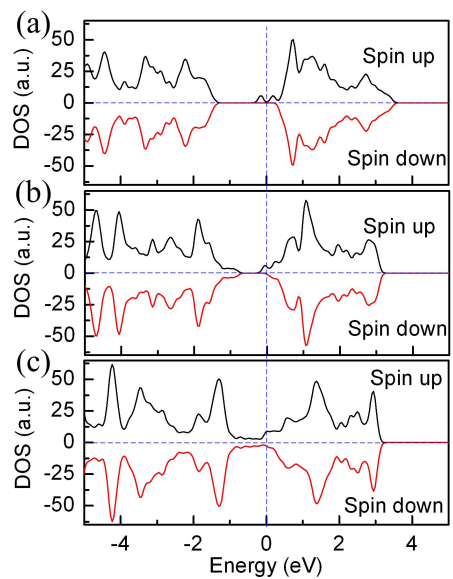

Figure 3. Calculated spin polarized density of states (DOS) of NO+MoS$_2$ under strain of (a) 0%, (b) 5%, and (c) 12.5%.

**Adsorption of $NH_3$ on monolayer $MoS_2$**  From the results in Table 1, one can see that adsorption of ammonia on monolayer $MoS_2$ also exhibits an appreciable response to applied strain. We have performed extensive calculations in the strain range of 0~12.5% to explore the behavior of ammonia adsorption on strained $MoS_2$. The calculated adsorption distance and energy as a function of strain are shown in Fig. 4. Here the results follow the same general trend for NO adsorption shown in Fig. 2a. There are, however, notable differences in the variation of these quantities in response to increasing strain. The decrease of the adsorption distance and energy is gentler before the metallization of the $MoS_2$ layer, and the curves form a plateau near the critical strain region (around 10% where $MoS_2$ is metalized)[14-17]. These results give a quantitative description of the relatively weaker strain dependence of ammonia adsorption on $MoS_2$. Further increasing strain, however, leads to a very sharp reduction of the adsorption distance with greatly increased adsorption energy (see Fig. 4a). A detailed structural examination shows that the $MoS_2$ layer is broken along its zigzag direction (see the inset of Fig. 4a for a snapshot at 12.5% strain), and the crack line is located at the ammonia adsorption site. This result reveals a strong corrosion effect of ammonia gas on the mechanical property of the $MoS_2$ layer. Here the breaking critical strain of 12.5% for $NH_3$ adsorbed $MoS_2$ monolayer is significantly smaller than previous theoretical prediction of 20% for pristine $MoS_2$ monolayer. The present result is closer to the recent experimentally measured value of about 10%. It shows that ammonia molecules have a detrimental effect on degrading the mechanical integrity of the $MoS_2$ layer, and they should be avoided, perhaps through proper chemical functionalization, in applications of $MoS_2$ as components in flexible electronics.

Similar to the situation of NO adsorption, the enhanced adsorption characteristics of ammonia on $MoS_2$ also results from increased charge transfer under strain. From Fig. 4(b) to (d), one can see a systematically enhanced charge transfer with increasing strain, going from 0.083e at 0% strain to 0.13e at 12.5% strain according to a Mulliken charge analysis. As previously reported[22] and also observed in the present work, the NH$_3$ adsorption has little effect on the electronic structure of MoS$_2$. The electronic band gap of NH$_3$ adsorbed strain-free monolayer MoS$_2$ is 1.5 eV, which is identical to the gap of pristine strain-free monolayer MoS$_2$. This is because the adsorption induced states are located in the high-lying conduction bands in the energy range of 4~6 eV, and they have little effect on the states near the Fermi level. Under strain, however, charge transfer significantly impacts the electronic structure, reducing the gap to 0.32 eV at 5% strain and closing it at 10% strain as indicated by the DOS plots shown in Fig. 4e.

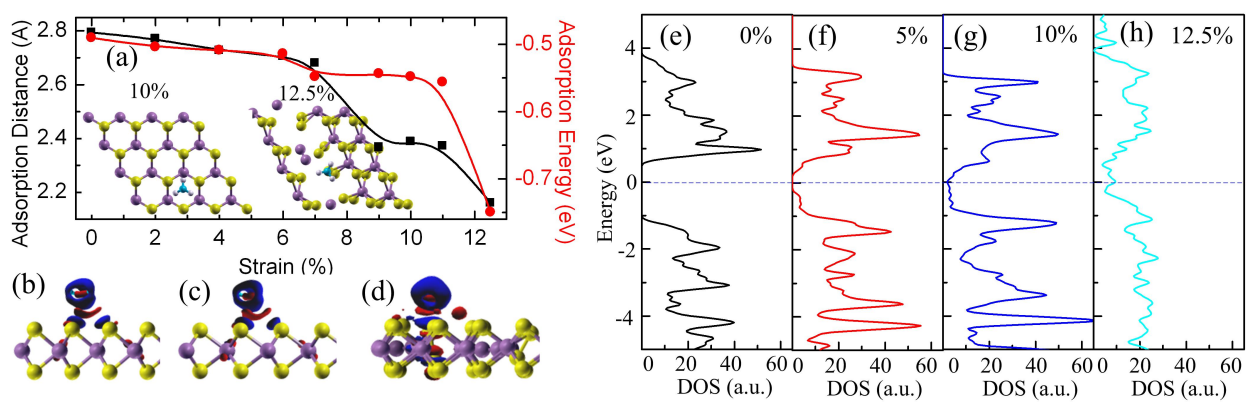

Figure 4. (a) Calculated adsorption distance and energy as a function of strain for NH$_3$ adsorbed monolayer MoS$_2$. The structural snapshots at selected strained conditions are presented as insets. Charge transfer is plotted for the strain state of (b) 0%; (c) 10% and (d) 12.5%. (e-h) DOS of NH$_3$+MoS$_2$ under different strained conditions.

**Discussion** We have discovered by ab initio calculations a highly sensitive strain dependence of adsorption of NO and NH$_3$ molecules on monolayer MoS$_2$. The adsorption strength as measured by the binding energy of these molecules to the MoS$_2$ layer can be controlled and enhanced by applied strain, and it leads to significantly changed electronic and magnetic properties, which can be used as ultrasensitive markers to detect these chemical species. The tunable magnetic moment of the adsorbed NO molecules and the accompanied transition to a half-metallic state may also find application in nanoscale electronics and spintronics. Our results reveal that these new findings are driven by the enhanced electron transfer between the adsorbates and the $MoS_2$ layer, especially after the strain-induced metallization of the $MoS_2$ layer. This mechanism is also responsible for the considerable deterioration of mechanical strength of $NH_3$ adsorbed monolayer $MoS_2$, which explains the large discrepancy between recent experimental results and previous theoretical predictions and suggests possible remedies for improving the mechanical properties that are important for the application of $MoS_2$ films in electromechanical devices. Since the adsorption of NO and $NH_3$ on strain-free $MoS_2$ layer and the manipulation of strained $MoS_2$ films have been recently demonstrated by separate experiments, we expect that our predicted strain engineering of selective chemical adsorption on $MoS_2$ would stimulate immediate interest and further exploration for fundamental understanding and practical application.

**Methods**

The reported calculations were performed using the density functional theory (DFT) approach based on pseudopotentials with localized atomic-orbital basis sets as implemented in the SIESTA code.[26] We used the Perdew−Burke−Ernzerhof (PBE)[27] generalized gradient approximation (GGA) for the exchange-correlation energy and norm-conserving pseudopotentials for the core−valence interactions. Spin polarization was included when calculating the adsorption of NO and $NO_2$ on $MoS_2$ since these molecules are paramagnetic, but not considered in the calculations for other gas molecules. The double-$\zeta$ polarized numerical atomic-orbital basis sets for all atoms were used. Single k-point and 5 × 5 × 1 Monkhorst-Pack k-point grid were used in the structural optimizations and energy calculations, respectively. The plane-wave cutoff of 300 Hartree was chosen for all simulations, and atomic positions were fully relaxed until the force on each atom <0.02 eV/Å. To isolate the nanosheets, the supercells contain a vacuum layer of at least 10 Å along the vertical direction to guarantee negligible interactions between the neighboring layers. The lattice constant of strain-free $MoS_2$ unit cell is taken as the experimental value, $a_0$=3.166 Å,[28] and the strained nanosheet takes the lattice constant of $a=(1+\varepsilon)\times a_0$.

**Acknowledgements**

L.K. acknowledges financial support by the Alexander von Humboldt Foundation of Germany. C.F.C. was supported by the Department of Energy Cooperative Agreement DE-NA0001982.

**References**

1. Wang, Q. H.; Kourosh Kalantar-Zadeh, K.; Kis, A.; Coleman, J. N. and Michael Strano, S. Electronics and optoelectronics of two-dimensional transition metal dichalcogenides, Nature Nanotech. **2012**, 7, 699–712.

2. Radisavljevic, B.; Radenovic, A.; Brivio, J.; Giacometti V. and Kis, A. Single-layer $MoS_2$ transistors, Nature Nanotech. **2011,** 6, 147–150.

3. Popov, I.; Seifert, G. and Tomanek, D. Designing Electrical Contacts to $MoS_2$ Monolayers: A Computational Study, Phys. Rev. Lett. **2012**, 108, 156802.

4. Butler, S. Z.; et al, Progress, Challenges, and Opportunities in Two-Dimensional Materials Beyond Graphene, ACS Nano **2013**, 7, 2898-2926.

5. Perkins, F. K.; Friedman, A. L.; Cobas, E.; Campbell, P. M.; Jernigan G. G. and Jonker, B. T. Chemical Vapor Sensing with Monolayer $MoS_2$, *Nano Lett.*, **2013**, 13, 668–673

6. Zhang, W.; Huang, J.-K.; Chen, C.-H.; Chang, Y.-H.; Cheng, Y.-J.; Li, L.-J.; High-Gain Phototransistors Based on a CVD $MoS_2$ Monolayer, Adv. Mater. **2013**, 25, 3456–3461.

7. Li, H.; Yin, Z.; He, Q.; Li, H.; Huang, X.; Lu, G.; Fam, D. W. H.; Tok, A. I. Y.; Zhang, Q.; and Zhang, H. Fabrication of Single- and Multilayer $MoS_2$ Film-Based Field-Effect Transistors for Sensing NO at Room Temperature, Small **2012**, 8, 63-67.

8. He, Q.; Zeng, Z.; Yin, Z.; Li, H.; Wu, S.; Huang, X. and Zhang, H. Fabrication of Flexible $MoS_2$ Thin-Film Transistor Arrays for Practical Gas-Sensing Applications, Small **2012**, 8, 2994-2999.

9. Late, D. J.; et al, Sensing Behavior of Atomically Thin-Layered $MoS_2$ Transistors, *ACS Nano*, 2013, 7, 4879–4891.

10. Lee, K.; Gatensby, R.; McEvoy, N. Hallam, T. and Duesberg, G. S. High Performance Sensors Based on Molybdenum Disulfide Thin Films, Adv. Mater. **2013**, DOI: 10.1002/adma.201303230.

11. Yao, Y.; Tolentino, L.; Yang, Z.; Song, X.; Zhang, W.; Chen, Y. and Wong, C.-P. High-Concentration Aqueous Dispersions of MoS$_2$, Adv. Funct. Mater. **2013**, 23, 3577–3583.

12. Sun, Q.; Li, Z.; Searles, D. J.; Chen, Y.; Lu, G. and Du, A. Charge-Controlled Switchable CO$_2$ Capture on Boron Nitride Nanomaterials, J. Am. Chem. Soc. **2013**, 135, 8246−8253.

13. Bissett, M. A.; Konabe, S.; Okada, S.; Tsuji, M. and Ago, H. Enhanced Chemical Reactivity of Graphene Induced by Mechanical Strain, ACS Nano, **2013**, DOI: 10.1021/nn404746h.

14. Bhattacharyya, S.; Singh, A. K. Semiconductor–Metal Transition in Semiconducting Bilayer Sheets of Transition-Metal Dichalcogenides, Phys. Rev. B **2012**, 86, 075454.

15. Scalise, E.; Houssa, M.; Pourtois, G.; Afanas'ev, V. and Stesmans, A. Strain-Induced Semiconductor to Metal Transition in the Two-Dimensional Honeycomb Structure of MoS$_2$, Nano Res. **2012**, 5, 43–48

16. Yun, W. S.; Han, S. W.; Hong, S. C.; Kim, I. G.; Lee, J. D. Thickness and Strain Effects on Electronic Structures of Transition Metal Dichalcogenides: 2H-MX$_2$ Semiconductors (M = Mo, W; X = S, Se, Te) *Phys. Rev. B* **2012**, 85, 033305

17. Johari, P. and Shenoy, V. B. Tuning the Electronic Properties of Semiconducting Transition Metal Dichalcogenides by Applying Mechanical Strains, *ACS Nano*, **2012**, 6, 5449–5456.

18. Bertolazzi, S.; Brivio, J. and Kis, A. Stretching and Breaking of Ultrathin MoS$_2$, ACS Nano, **2011**, 5, 9703-9709.

19. Castellanos-Gomez, A.; Poot, M.; Steele, G. A.; van der Zant, H. S. J.; Agraït, N. and Rubio-Bollinger, G. Elastic Properties of Freely Suspended MoS$_2$ Nanosheets, Adv. Mater. **2012**, 24, 772–775.

20. Castellanos-Gomez, A.; Poot, M.; Steele, G. A.; van der Zant, H. S. J.; Agraït, N. and Rubio-Bollinger, G. Mechanical properties of freely suspended semiconducting graphene-like layers based on MoS$_2$, Nanoscale Res. Lett. **2012**, 7, 233.

21. Li, T. Ideal strength and phonon instability in single-layer MoS$_2$, Phys. Rev. B, **2012**, 85, 235407.

22. Peng, Q. and De, S. Outstanding mechanical properties of monolayer MoS$_2$ and its application in elastic energy storage, Phys. Chem. Chem. Phys. **2013**, DOI: 10.1039/c3cp52879k

23. Zhao, S.; Xue, J. and Kang, W. Gas Absorption on MoS$_2$ Monolayer from First-principles Calculations, arXiv:1306.6751, 2013.

24. Kresse, G. and Furthmüller, J. Efficient iterative schemes for ab initio total-energy calculations using a plane-wave basis set. Phys. Rev. B, **1996**, 54, 11169.

25. Leenaerts, O.; Partoens, B. and Peeters, F. M. Adsorption of H$_2$O, NH$_3$, CO, NO$_2$, and NO on graphene: A first-principles study, Phys. Rev. B **2008**, 77, 125416.

26. Soler, J. M.; Artacho, E.; Gale, J. D.; García, A.; Junquera, J.; Ordejón, P.; Sánchez-Portal, D. The SIESTA Method for Ab Initio Order-N Materials Simulation, J. Phys.: Condens. Matter **2002**, 14, 2745

27. Perdew, J. P.; Burke, K.; Ernzerhof, M. Generalized Gradient Approximation Made Simple, Phys. Rev. Lett. **1996**, 77, 3865

28. Powell, A. V.; Kosidowski, L. and McDowall, A. Inorganic–organic hybrids by exfoliation of MoS$_2$, J. Mater. Chem., **2001**, 11, 1086-1091.